%% file: QHGrphv10.tex
\documentclass[aps,floats,twocolumn,prb,10pt]{revtex4-2}
\usepackage[unicode,linktocpage=true]{hyperref}
\hypersetup{colorlinks=true,linkcolor=blue,urlcolor =blue,citecolor=blue,anchorcolor=blue}
\usepackage{amsmath}
\usepackage{graphicx}
\usepackage{xcolor}

\def\nn{\nonumber}

\begin{document}

\title{Dipole representation of composite fermions in graphene quantum Hall systems}

\author{Sonja Predin}
\email{sonja@ipb.ac.rs}
\affiliation{Scientific Computing Laboratory, Center for the Study of Complex Systems,Institute of Physics Belgrade, University of Belgrade, Pregrevica 118, 11080 Belgrade, Serbia}

\begin{abstract}
The even denominator fractional quantum Hall effect has been experimentally observed 
in graphene in the fourth Landau level ($\mathcal{N} = 3$). 
This paper is motivated by recent studies regarding the possibility of pairing and the nature of the 
ground state in this system. By extending the dipole 
representation of composite fermions, we adapt this framework to the context of graphene’s quantum 
Hall systems, with a focus on half-filled Landau levels. 
We derive an effective Hamiltonian that incorporates the 
key symmetry of half-filled Landau levels, particularly particle-hole symmetry. At the Fermi level, the 
energetic instability of the dipole state is influenced by the interplay between topology and symmetry, 
driving the system towards a critical state. We explore the possibility that this critical state 
stabilizes into one of the paired states with well-defined pairing solutions. However, our results 
demonstrate that the regularized state which satisfies boost invariance at Fermi level and lacks well-
defined pairing instabilities emerges as energetically more favorable. Therefore, we find no well-
defined pairing instabilities of composite fermions in the dipole representation in the half-filled 
fourth Landau level ($\mathcal{N} = 3 $) of electrons in graphene. Although the theory of composite fermions has 
its limitations, further research is required to investigate other possible configurations.  
We discuss the consistency of our results with experimental and numerical studies, and their relevance 
for future research efforts. 
\end{abstract}

\maketitle

\section{Introduction} 
\label{intro}

The discovery of the fractional quantum Hall effect (FQHE) \cite{tsui1982} represents a significant 
breakthrough in the study of topological states of matter, introducing a rich landscape of quantum 
phases driven by electron correlations in two-dimensional electron systems (2DES) \cite{girvin2019}. 
Among the most intriguing manifestations 
of the FQHE is the observation of a quantized Hall plateau at the filling factor $\nu = 5/2$ 
\cite{willett}, indicating the FQHE at the half-filled second Landau level ($ n = 1 $). 
The observation of 
even-denominator FQHE states is astonishing, considering that the fermionic statistics of electrons 
suggest the denominator of the fraction should be an odd number. These significant discoveries have 
inspired intensive theoretical and experimental efforts to decode the underlying physics and 
implications of these states.

The concept of composite fermions (CFs), introduced by Jain \cite{jain1989a, jain2007}, has provided an 
insightful framework for comprehending various aspects of quantum Hall phenomena, primarily for the 
half-filled second LL ($ n = 1 $). CFs are quasiparticles formed by attaching an even number of quantized vortices 
to electrons.

In recent years, advancements in noncommutative field theory in high-energy physics, have inspired Dong 
and Senthil \cite{dong2020} to revisit longstanding problems in quantum Hall physics. In particular, 
this includes the problem of the LL at $\nu = 1$ for bosons, initially formulated by Pasquier 
and Haldane \cite{pasquier}, and further developed by Read \cite{read}. This 
problem involves an additional degree of freedom, namely the vortex degree of freedom, 
where each boson is 
associated with a single flux of the magnetic field, and since vortices are fermionic, and the resulting 
composite quasiparticles are neutral, resembles CFs. Furthermore, CFs experience a zero average 
effective magnetic field, similar to electrons at $\nu = 1/2$. Unlike previous approaches that relied on 
an averaged field energy, Dong and Senthil \cite{dong2020} introduced the concept of intrinsic dipole 
energy. The application of noncommutative field theory to quantum Hall systems has since led to 
significant advancements, with recent studies extending these concepts to various quantum Hall states 
\cite{predin2021, ma2022, dong2022, goldman2022, predin2023a, predin2023b}.

Moreover, fractional quantum Hall (FQH) states with additional half-integer filling factors have been observed in various materials, including graphene—a monolayer of carbon atoms arranged in a hexagonal lattice. Graphene’s unique electronic and topological properties make it an ideal platform for studying exotic quantum phenomena, including the fractional quantum Hall effect (FQHE). Notably, a recent experimental study by Kim et al. \cite{kim2019} identified FQHE states at half-filling in the fourth Landau level (\(\mathcal{N} = 3\)) of monolayer graphene.

Additionally, research by Sharma et al. \cite{balram2023} has investigated CF pairing in monolayer graphene, using alternative pairing functions and numerical simulations on a torus within the Bardeen-Cooper-Schrieffer (BCS) framework. The microscopic CF-BCS theory has been highly successful in capturing many known pairing instabilities, particularly in two-dimensional electron systems (2DES) such as GaAs at filling factors \( \nu = 1/2 \) and \( \nu = 5/2 \) \cite{sharma}. Remarkably, the CF-BCS approach has also effectively described instabilities in wide quantum wells at filling factors \( \nu = 1/2 \) and \( \nu = 1/4 \) \cite{sharma2024}. In the case of graphene, the authors in Ref. \cite{balram2023} concluded that the BCS variational state for CFs reveals an $f$-wave pairing instability in the \(\mathcal{N} = 3\) LL.

Furthermore, they suggested the possibility of a $p$-wave instability for the 
$\mathcal{N} = 2$ LL, where FQHE has not been experimentally observed. However, these findings are in contrast 
with numerical studies on the sphere, which have not observed such pairing instabilities \cite{kim2019}. 
The disparity between these results highlights the need for further investigation. 
Besides, topological pairing, particularly $p$-wave and $f$-wave, is crucial 
for quantum computing applications due to its non-Abelian statistics, which enable robust qubits and 
fault-tolerant quantum operations \cite{kitaev}.

The dipole representation of FQHE states, introduced in Ref. \cite{predin2023a}, is particularly 
relevant to this study. In this representation, dipoles are neutral composite objects formed by an 
electron and its correlation hole. These dipoles possess moments proportional to their momentum in an 
external magnetic field. This effective theory has proven to be robust, capturing the microscopic 
description of several problems and aligning with previous experimental and numerical studies. In 
particular, it yielded results consistent with numerical and experimental studies of 2DES, where the 
Pomeranchuk instability is observed in higher half-filled LLs. A recent study \cite{nikola} even 
provided an explanation for the mechanism of $p$-wave pairing of CFs at half-filling in the second LL 
($ n = 1 $) of 
electrons in 2DES and in the fully filled first LL ($ n = 0 $) of bosons, further underscoring the importance of 
this representation.

In this paper, we aim to explore the potential for CF pairing in 
monolayer graphene using the BCS framework, also utilizing the dipole 
representation. The half-filled LL system possesses an additional feature, the particle-hole (PH) 
symmetry \cite{predin2023a}, which implies that the density of holes corresponds to the density 
of composite holes. Consequently, this state is energetically unstable with respect to repulsive 
interactions, leading to a critical state. 
Through analytical studies, we demonstrate that in the fourth 
LL ($\mathcal{N} = 3$) of a graphene monolayer, this critical state is not stabilized by selecting one of the 
two possible symmetry-broken paired states; instead, it stabilizes into a regularized non-paired state, 
which cannot support gapped pairing solutions due to the absence of mass. Our findings are consistent 
with numerical studies conducted on spherical geometries \cite{kim2019}.  

This paper is organized as follows: In Section \ref{pasquier-haldane}, we introduce the necessary 
formalism and key concepts of the dipole representation and apply them to the half-filled LLs of 
electrons in graphene. In particular, we discuss the model, and the effective Hamiltonian in the 
dipole representation. Section 
\ref{paired} explores the possibility and mechanism of paired states within the context of the dipole 
representation and analyzes their stability. Finally, in Section \ref{conclusion}, we conclude with a 
discussion of our findings and their implications for future research. 

\section{Implementing Boost Invariance and State Regularization in Graphene's Landau Levels}
\label{pasquier-haldane}

\subsection{The model}
\label{model}

Graphene exhibits unique electronic properties, including a "relativistic" quantum Hall effect due 
to its charge carriers, which behave as massless Dirac fermions \cite{novoselov, 
zhang2005}. In graphene, electrons are arranged in a honeycomb lattice composed of two sublattices, $A$ 
and $B$, and their behavior can be described by a two-dimensional Dirac equation \cite{neto}. In a 
tight-binding model, we consider only the nearest neighbor hopping parameter $t$ between sites $A$ and 
$B$. The low energy properties of graphene are captured by a two-band model labeled as $\lambda = \pm$ 
where the dispersion is linear. The valence band $\lambda = +$ and the conduction band $\lambda = -$ 
touch each other at the two inequivalent corners $K_+$ and $K_-$ (referred to as valleys) of the 
Brillouin zone \cite{neto, predin2016}.  

When an external magnetic field is applied, the Hamiltonian for low-energy states around the $K_+$ valley is given by \cite{divincenzo, mccann}:
\begin{equation}
H = v \begin{pmatrix}
0 & \pi^{\dagger} \\
\pi & 0.
\end{pmatrix}
\label{ham_mag}
\end{equation}
where $v = \frac{\sqrt{3}}{2} \frac{at}{\hbar}$ is the velocity, and the operators $
\pi^\dagger$ and $\pi$, in the Landau gauge, coincide with the LL creation and annihilation operators, 
respectively \cite{mccann}.

Here, we focus on spinless electrons confined to the $\mathcal{N}$-th LL, where only 
intra-LL excitations are considered.

The spinor states in the $\lambda-$band, obtained from the two-dimensional Dirac equation, can be expressed as:
\begin{equation}
\psi_{\lambda \mathcal{N} ,m}^{\xi} = \frac{1}{\sqrt{2}} \begin{pmatrix}
 \vert \, \mathcal{N}-1, \, m \, \rangle \\
\lambda \vert \, \mathcal{N}, \, m \, \rangle
\end{pmatrix},
\label{spinor1}
\end{equation}
for $\mathcal{N} \neq 0$
\begin{equation}
\psi_{\mathcal{N}=0,m}^{\xi} = \begin{pmatrix}
0 \\
\vert 0, \, m \, \rangle
\end{pmatrix},
\label{spinor2}
\end{equation}
for $\mathcal{N} = 0$, in terms of the harmonic oscillator states $\left\vert \, \mathcal{N}, \, m \, \right\rangle$ and the 
guiding-center quantum number $m$. These spinors describe states within the $ \mathcal{N} $-th LL in the band $
\lambda$. The first component of the spinor represents the amplitude on the $A$ sublattice at the point 
$K_+$ ($\xi = +$) and the amplitude on the $B$ sublattice at the point $K_-$ ($\xi = -$).

\subsection{The Pasquier-Haldane-Read construction for half-filled LL of electrons in graphene}
\label{phr}

In this section, we extend the traditional Pasquier-Haldane-Read (PHR) construction \cite{pasquier, 
read} to describe the complex excitations in half-filled LLs of electrons in graphene. The Pasquier-Haldane 
approach reformulates the theory by using an expanded Hilbert space of composite fermions, rather than 
the original Hilbert space of bosons, to represent the GMP algebra. Unlike the 
original model, which was formulated for fully filled LLs of bosons, our approach adapts the PHR 
construction to systems where LLs are half-filled of electrons. 
In this paper, we employ an expansion of the Hilbert 
space by incorporating correlation holes, which are positive charges that pair with electrons to form 
neutral dipoles. This representation is crucial for describing systems with PH symmetry, a key 
characteristic of half-filled LLs, and for maintaining the topological properties of the quantum Hall 
states.

We begin by representing the basis states in a LL as:
\begin{equation}
\{|o_1\rangle, |o_2\rangle, \ldots, |o_{N_\phi/2}\rangle, |m_1\rangle, |m_2\rangle, \ldots, |m_{N_\phi/2}\rangle\},
\end{equation}
where $|o_i\rangle$ and $|m_i\rangle$ represent the states, which form the foundation for describing the system in our enlarged Hilbert space, which accommodates both electrons and correlation holes \cite{predin2023a}. Here, $ N_\phi $ represents the number of orbitals.

In this expanded space, we define creation and annihilation operators that satisfy the algebra:
\begin{eqnarray}
&\lbrace c_{mo}, c_{o'm'}^\dagger \rbrace = \delta_{mm'} \delta_{oo'}, \nn \\
&\lbrace c_{mo}, c_{o'm'} \rbrace = 0, \nn \\
&\lbrace c_{mo}^{\dagger}, c_{o'm'}^\dagger \rbrace = 0.
\label{comm}
\end{eqnarray}  

CFs operators $ c_{mo} $ and $ c^{\dagger}_{mo} $ in the LL have double indices, where m and n labels physical electrons and correlation holes, resprectively.
Using these operators, we define the Fourier components of the physical (left) and unphysical (right) densities in the $\mathcal{N}$-th LL, incorporating the form factor $ F_{\mathcal{N}}(\vert \vec{q} \vert) $ :
\begin{align}
&\rho_{\mathcal{N}}^{L}(\vec{q})=F_{\mathcal{N}}(\vert \vec{q} \vert) \sum_{o,\, o'} \sum_{m} \langle o \vert \tau_{-\vec{q}} \vert o' \rangle c_{mo}^{\dagger}c_{o'm},
\label{ro1} \\
&\rho_{\mathcal{N}}^{R}(\vec{q})=F_{\mathcal{N}}(\vert \vec{q} \vert) \sum_{m,\, m'} \sum_{o} \langle m \vert \tau_{-\vec{q}} \vert m' \rangle c_{mo}^{\dagger}c_{om'}.
\label{ro2}
\end{align}

Here, $ \tau_{\vec{q}}=e^{i\vec{q}\vec{R}} $ is the translation operator, 
where $ \vec{R} = \left( X, Y\right) $ represents the guiding center coordinates of a CF \cite{prange, goerbig, shankar2001} in the external magnetic field $ \vec{B} = - B \vec{e}_{z} $, with components obeying the commutation relation:
\begin{equation}
\left[X, Y\right]= i l_B^2.
\label{commutation}
\end{equation}
Here, $ l_B $ is the magnetic length. In what follows, we will set $ l_B \equiv 1 $. 
This framework ensures that the neutral CFs are accurately described within the context of the LL dynamics.

The Fourier components of the form factor in the $\mathcal{N}$-th LL are given in 
terms of Laguerre polynomials $ L_{\mathcal{N}} $ in the following way:
\begin{equation}
F^{(\mathcal{N})}(q) = 
\begin{cases}  1 & \text{if } \mathcal{N} = 0, \\
\frac{1}{2} \left[ L_{\mathcal{N}-1} \left( \frac{q^2}{2} \right) + L_{\mathcal{N}} \left( \frac{q^2}{2} \right) \right] & \text{if } \mathcal{N} \neq 0.
\end{cases}
\end{equation}
For $ \mathcal{N} = 0 $, the form factor for graphene coincides with that of a 2DES, such as GaAs.
However, for higher $ \mathcal{N} $ it can be viewed as an average of the two form factors of those for 2DES \cite{goerbig2004}.  

Furthermore, the annihilation operator can be written in relation to its momentum space representation in the following way:
\begin{eqnarray}
c_{mo}=\int \frac{d\vec{k}}{(2\pi)^{2}}\langle m \vert \tau_{\vec{k}} \vert o \rangle c_{\vec{k}}.
\label{CF_operator}
\end{eqnarray}

By substituting Eq.~(\ref{CF_operator}) into the equations for the left and right density operators, 
Eq.~(\ref{ro1}) and Eq.~(\ref{ro2}), we obtain:
\begin{align}
&\rho_{\mathcal{N}}^L(\vec{q}) = F_{\mathcal{N}}(\vec{q}) \int \frac{d^2\vec{k}}{(2\pi)^2} e^{\frac{i}{2}\vec{k}\times \vec{q}}c_{\vec{k}-\vec{q}}^\dagger c_{\vec{k}} , \label{rho3} \\
&\rho_{\mathcal{N}}^R(\vec{q}) = F_{\mathcal{N}}(\vec{q}) \int \frac{d^2\vec{k}}{(2\pi)^2} e^{-\frac{i}{2}\vec{k}\times \vec{q}}c_{\vec{k}-\vec{q}}^\dagger c_{\vec{k}}. \label{rho4}
\end{align}

Using the relations Eq.~(\ref{comm}), it is easy to show that these densities obey the Girvin-MacDonald-Platzman (GMP) algebra \cite{GMP}:
\begin{align}
&\left[\rho_{0}^{L}(\vec{q}), \rho_{0}^L(\vec{q}')\right] = 2i\sin\left(\frac{\vec{q}\times \vec{q}'}{2}\right) \rho_{0}^{L}(\vec{q}+\vec{q}'), \nn \\
&\left[\rho^R_{0}(\vec{q}), \rho^R_{0}(\vec{q}')\right] = - 2i\sin\left(\frac{\vec{q}\times \vec{q}'}{2}\right)\rho_{0}^{R}(\vec{q}+\vec{q}'), \nn \\
&\left[\rho^L_{0}(\vec{q}), \rho_{0}^R(\vec{q}')\right] = 0.
\label{commutation2}
\end{align}
This is induced by the commutation relation Eq.~(\ref{commutation}), and restriction to the single LL.
The realization of the GMP algebra using the canonical CF variables facilitates the application of mean-
field methods \cite{read}.

\subsection{The effective Hamiltonian in dipole represenation in graphene}

To accurately describe the system within the dipole representation, it is essential to impose a specific 
constraint. These constraints serve several purposes.
First, the constraint ensures that the number of degrees of freedom in the effective model aligns with 
the microscopic description of the  physical problem. Furthermore, the constraint also defines the 
physical subspace within the enlarged Hilbert space, which includes 
additional degrees of freedom such as correlation holes. Second, the imposed constraint must preserve 
the PH symmetry, a fundamental characteristic of the half-filled LL, ensuring that the model accurately 
reflects the physical properties of the system.

We introduce the following constraint:
\begin{equation}
\rho_{\mathcal{N} \mathcal{N}}^L + \rho_{\mathcal{N} \mathcal{N}}^R = 1
\label{constrain1}
\end{equation}

This constraint acts as a null operator in momentum space:
\begin{equation}
\rho_{\vec{q}}^L + \rho_{\vec{q}}^R = 0.
\label{constrain2}
\end{equation}

It is important to highlight that this constraint incorporates both physical and unphysical quantities, treating them as mutually dependent.
Since the right degrees of freedom (additional degrees of freedom) represent 
correlation holes in the enlarged space, this constraint effectively ensures, in the long-distance 
limit, that the density of correlation holes equals the density of real holes. 

Furthermore, when defining the problem in the enlarged space, operators, including the Hamiltonian, may
map physical states into superpositions of physical and unphysical states. To ensure 
that physical states remain within the physical subspace, the Hamiltonian must commute with the 
constraint.

The effective Hamiltonian in this framework must reflect the dipole representation within $ \mathcal{N} $-th LL. 
Additionally, it must preserve PH symmetry, meaning it remains invariant under the exchange of particles 
and holes. The Hamiltonian is carefully constructed to satisfy these conditions, and we impose a 
constraint that ensures the resulting Hamiltonian has a PH symmetric form \cite{predin2023a}

\footnotesize
\begin{equation}
\mathcal{H}_{\text{eff}} = \frac{1}{8} \int \frac{d^2 \vec{q}}{(2\pi)^2} V^{(\mathcal{N})}(\vec{q}) \left(\rho_0^{L}(\vec{q}) - \rho_0^{R}(\vec{q})\right)\left(\rho_0^{L}(-\vec{q}) - \rho_0^{R}(-\vec{q})\right).
\label{ham1} 
\end{equation}
\normalsize
Here, $ V^{(\mathcal{N})}(\vec{q}) = \frac{2 \pi e^2}{\epsilon|\vec{q}|} e^{-q^2/2} \left(F^{(\mathcal{N})}\right)^2 $ represents the effective interaction, which takes into 
account states within the $\mathcal{N}$-th LL. The form factor mimics the LL characteristics. Moreover, it should be noted that $ \upsilon~(\vec{q}) = \frac{2 \pi e^2}{\epsilon|\vec{q}|} $ defines the Fourier transform of the Coulomb interaction potential.

Using mean-field approximation and Hartree-Fock (HF) calculations, one easily finds that dispersion relation for graphene for the $\mathcal{N}$-th LL has the following form:
\begin{equation}
\varepsilon^{(\mathcal{N})}(\vec{k}) = \varepsilon_{0}^{(\mathcal{N})}(\vec{k}) + \varepsilon_{HF}^{(\mathcal{N})}(\vec{k}),
\label{energyTotal}
\end{equation}
where:
\begin{equation}
\varepsilon_{0}^{(\mathcal{N})}(\vec{k}) = \frac{1}{2}\int \frac{d^2 \vec{q}}{(2\pi)^2}V^{(\mathcal{N})}(\vert \vec{q} \vert)\sin^2\left(\frac{\vec{k}\times \vec{q}}{2}\right),
\label{energyTotal0}
\end{equation}
represents single-particle energy and
\begin{equation}
\varepsilon_{HF}^{(\mathcal{N})}(\vec{k}) = - \int \frac{d^2 \vec{q}}{(2\pi)^2}V^{(\mathcal{N})}(\vert\vec{k}-\vec{q}\vert)\sin^2\left(\frac{\vec{k}\times \vec{q}}{2}\right)n_{q},
\label{energyTotal0}
\end{equation}
represents the HF contributions. 
Here, $ n_q $ is the Fermi (step) function with $ n_{q} = 1 $ for $ q $ inside a circular Fermi surface
of radius $ q_F $, and zero otherwise.
This results aligns with Ref.~(\cite{shankar}), with the difference that here 
$ V(\vert \vec{q} \vert) $ represents the Coulomb interaction in the graphene system in the $\mathcal{N}$-th LL, and 
includes an additional factor of $ 1/4 $ that reduces the strength of the Coulomb interaction.
Furthermore, the single particle energy $ \varepsilon_{0}^{(\mathcal{N})} $ can be obtained in a closed analytical fashion as detailed in Appendix \ref{appendixA}. Moreover, in Appendix \ref{appendixA} we also give the complete data of
the corresponding energies for the lowest fourth LL ($ \mathcal{N} = 0, 1, 2, 3 $). On the other hand, we obtained the energy of interaction of this
particles $ \varepsilon_{HF}^{(\mathcal{N})} $ via numerical integration. 

The interaction of electrons within a single LL can be fully described by its Haldane pseudopotentials $ V_m $ \cite{haldane}, which quantify the interaction energies of two electrons with relative angular momentum $ m $. For the $ n $-th Landau level, the Haldane pseudopotentials are expressed as:  
\begin{equation}
V_m^{(n)} = \int \frac{d^2\vec{q}}{(2 \pi)^2} \, F^{(n)}(q) e^{-q^2} L_m(q^2),
\end{equation}  
where $ F^{(n)}(q) $ is the form factor associated with the $ n $-th LL.
In this study, we focus on two approximate effective interactions, $ V_{\text{Toke}} $ and $ V_{\text{Park}} $ \cite{toke, park}, which are defined in real space as:  
\begin{align}
V_{\text{Toke}}(r) &= \frac{1}{r} + \sum_{i = 0}^6 c_i\,r^i\, e^{-r}, \\
V_{\text{Park}}(r) &= \frac{1}{r} + a_1\, e^{-\alpha_1\,r^2} + a_2\, r^2\, e^{-\alpha_2\,r^2},
\end{align}  
respectively. It can be concluded that the effective interaction can be represented as the Coulomb interaction \( 1/r \) combined with short-range functions to account for deviations at short distances.
Thus, in addition to the Coulomb interaction, we will also utilize the Toke and Park interactions in the subsequent analysis. The coefficients $ c_i $ ($ i = 0, 1, 2, 3, 4, 5, 6 $) for $ V_{\text{Toke}} $ and $ a_1, a_2, \alpha_1, \alpha_2 $ for $ V_{\text{Park}} $ are determined by matching the effective interactions $ V_{\text{Toke}}(q) $ and $ V_{\text{Park}}(q) $, operating in the lowest LL (LLL), to the first seven (\( m = 0, 1, 2, 3, 4, 5, 6 \)) and four (\( m = 0, 1, 2, 3 \)) pseudopotentials of the Coulomb interaction \( V_{2m-1}^{(n)} \), respectively, in the second LL (\( n = 1 \)) and the fourth LL (\( n = 3 \)). The Fourier transforms of the effective interactions are obtained as:  
\begin{equation}
V_{\text{eff}}(q) = \int d^2 \vec{r} \, V_{\text{eff}}(r) e^{-i\vec{q}\cdot\vec{r}}.
\end{equation}  
The coefficients obtained through the symbolic solution of the corresponding system of equations for both interactions are listed in Table \ref{table_coeff}.

\begin{table}[ht!]
\centering
\begin{tabular}{|c|c|c|}
\hline
\textbf{Coefficient} & \textbf{$n = 1$} & \textbf{$n = 3$} \\ \hline
$c_0$ & -6.631 & 492.524 \\ \hline
$c_1$ & 13.298 & -976.021 \\ \hline
$c_2$ & -8.997 & 692.713 \\ \hline
$c_3$ & 2.934 & -235.342 \\ \hline
$c_4$ & -0.499 & 41.446 \\ \hline
$c_5$ & 0.0426 & -3.645 \\ \hline
$c_6$ & -0.00143 & 0.126 \\ \hline
$a_1$ & 0.0107017 & 11.8887 \\ \hline
$a_2$ & 0.109467 & -9.64883 \\ \hline
$\alpha_1$ & 0.038443 & 0.247147 \\ \hline
$\alpha_2$ & 0.446909 & 0.479972 \\ \hline
\end{tabular}
\caption{Values of coefficients $c_i$, $a_i$, $\alpha_i$ of the effective interactions for $n = 1$, and $n = 3$ LL in graphene.}
\label{table_coeff}
\end{table}

\subsection{Boost Invariance and State Regularization in Graphene}

In this section, we derive the Hamiltonian from the constraint previously established, ensuring that our 
effective theory maintains a valid microscopic description, at least at the Fermi level. As we noted above, the FQHE systems at half-filling in LLL exhibit the emergence of a Fermi liquid (FLL) state of composite quasiparticles (CFs for example). The Hamiltonian 
described in Eq.~(\ref{ham1}), which governs the dipole representation, incorporates a finite (bare) 
mass for single-particle energies as a consequence of the dipole structure. 
To achieve a well-defined Fermi 
liquid (FL) state description within the LL framework \cite{predin2023a}, it is essential for the 
Hamiltonian to exhibit boost invariance, which imposes a condition on the (bare) mass at the Fermi 
level.

To begin, we introduce an interaction term that is null within the physical space:
\footnotesize
\begin{equation}
\mathcal{H}_{0} = \mathcal{C}_{\mathcal{N}} \int \frac{d^2 \vec{q}}{(2\pi)^2} V^{(\mathcal{N})}(\vec{q}) \left(\rho_0^{L}(\vec{q}) + \rho_0^{R}(\vec{q})\right)\left(\rho_0^{L}(-\vec{q}) + \rho_0^{R}(-\vec{q})\right).
\label{ham2} 
\end{equation}
\normalsize
Therefore, the resulting Hamiltonian has the following form:
\begin{equation}
\mathcal{H}_{\text{res}} = \mathcal{H}_{\text{eff}} + \mathcal{H}_{\text{0}}.
\label{ham_res}
\end{equation}
Furthermore, we denote the constant $ \mathcal{C}_{\mathcal{N}} $ such that the energy $ \varepsilon_{1} $ denoting the single energy of the resulting Hamiltonian in Eq.~(\ref{ham_res}) fulfills the condition:
\begin{equation}
\frac{1}{m^*}=\frac{\partial^2 \varepsilon_{0}(\vec{k})}{\partial k^2} \Big|_{k = k_F} = 0,
\label{condition}
\end{equation}
where $ m^* $ is the effective mass. Therefore, the resulting Hamiltonian in Eq.~(\ref{ham_res}) has no terms $ \frac{k^2}{m^*} $, which contribute to the kinetic energy. The regularized state represents a FL-like (FLL) state, as is the case in the half-filled LLL of electrons.

\section{Dipole representation of composite fermions in graphene}
\label{paired}

\begin{figure*}[ht!]
\centering
\includegraphics[width=0.48\linewidth]{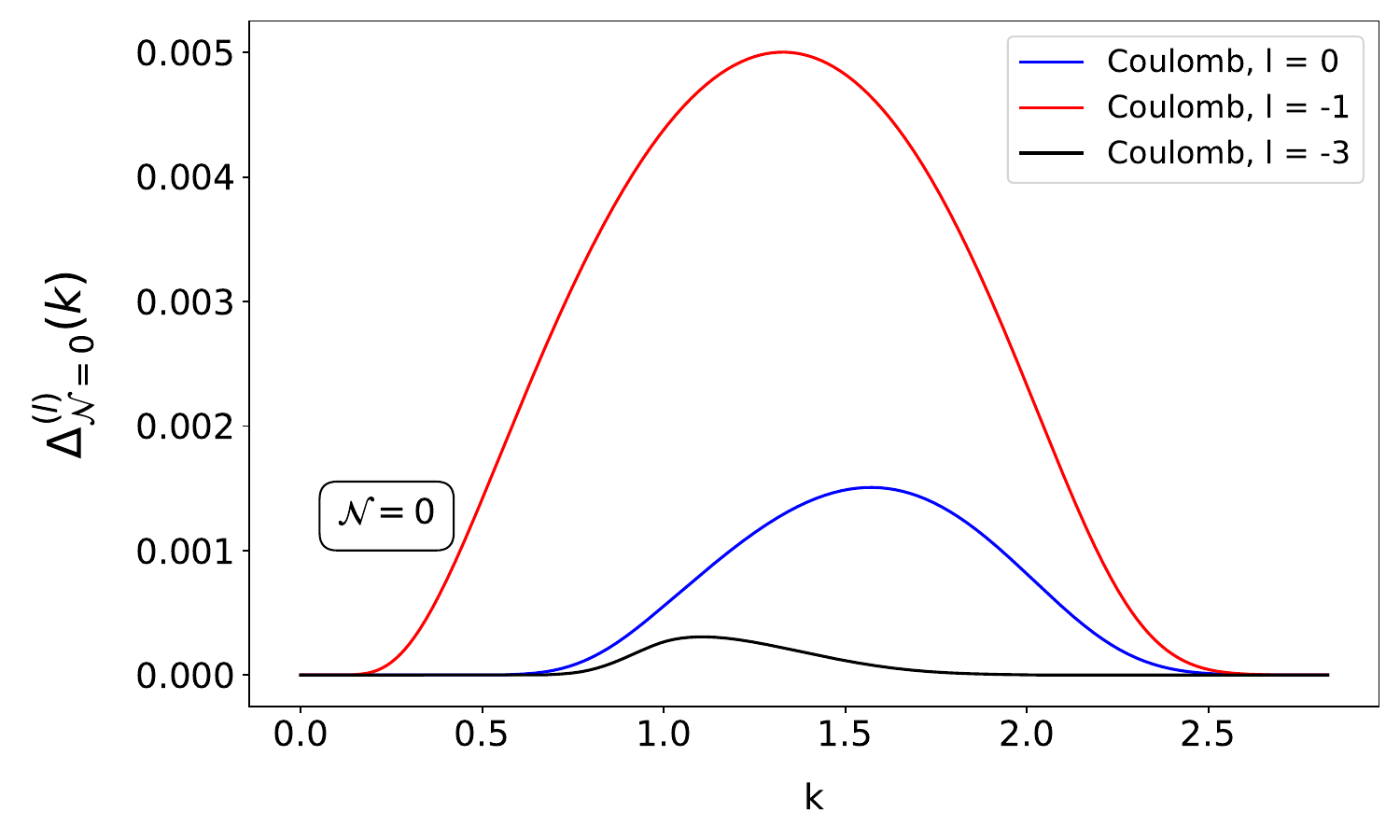}
\hfill
\includegraphics[width=0.48\linewidth]{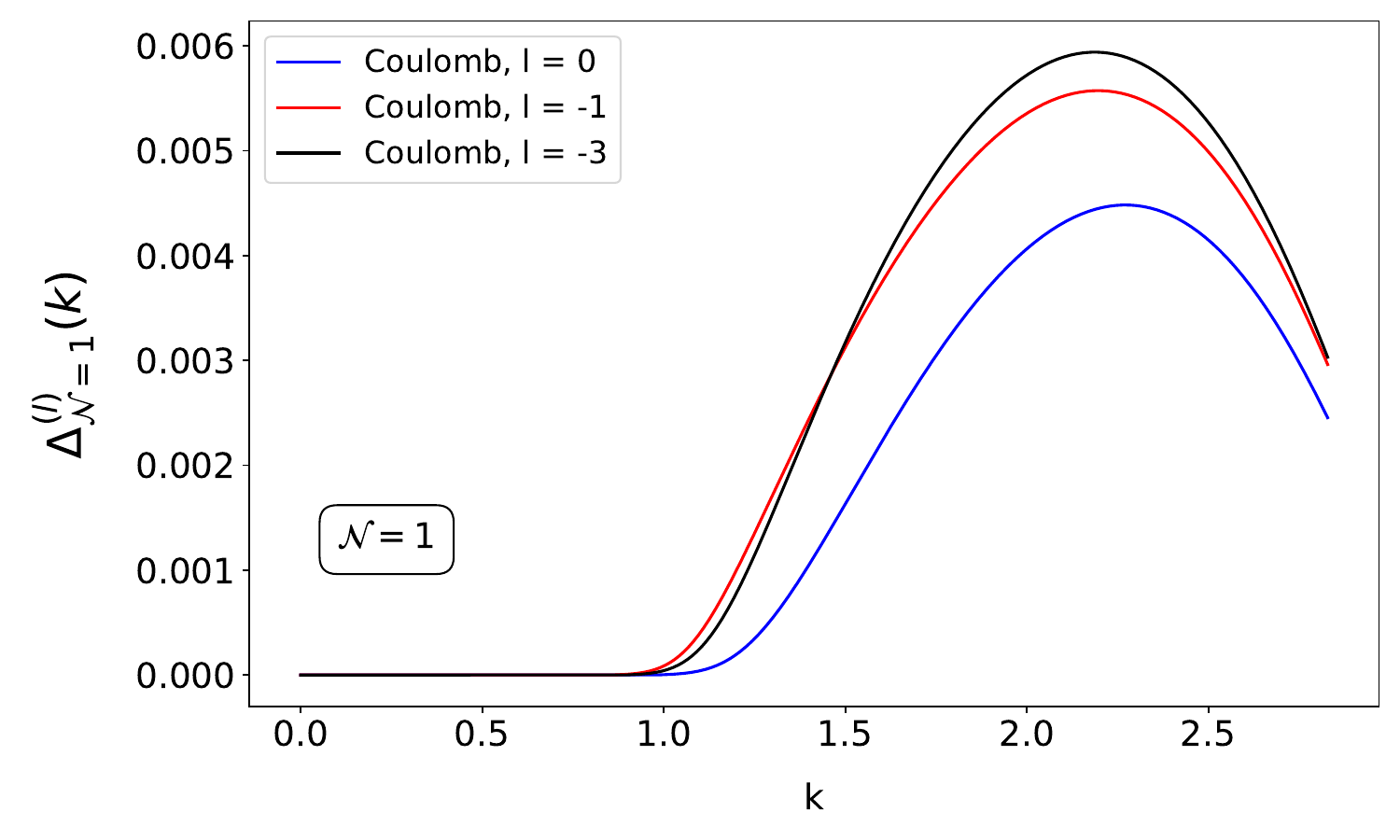}
\hfill
\includegraphics[width=0.48\linewidth]{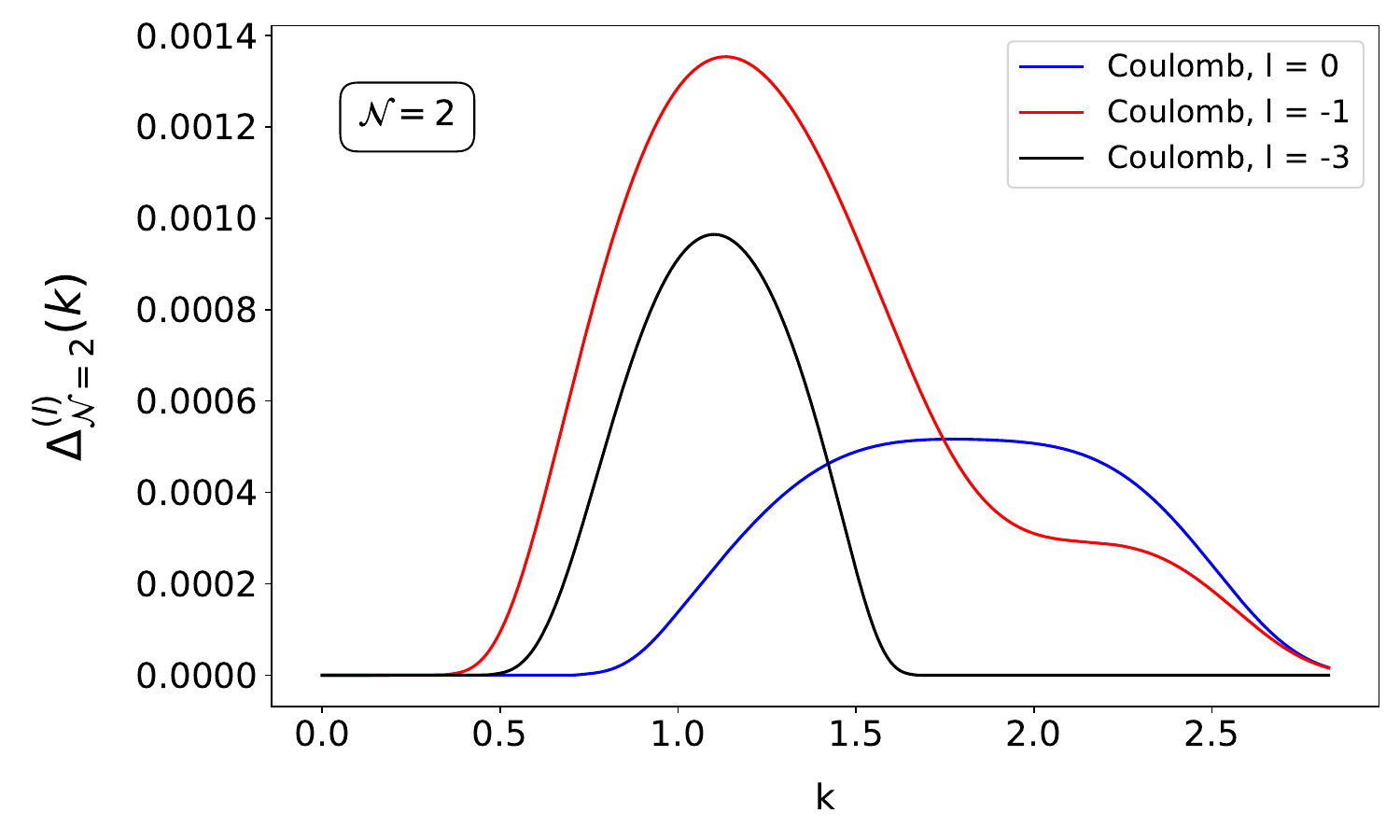}
\hfill
\includegraphics[width=0.48\linewidth]{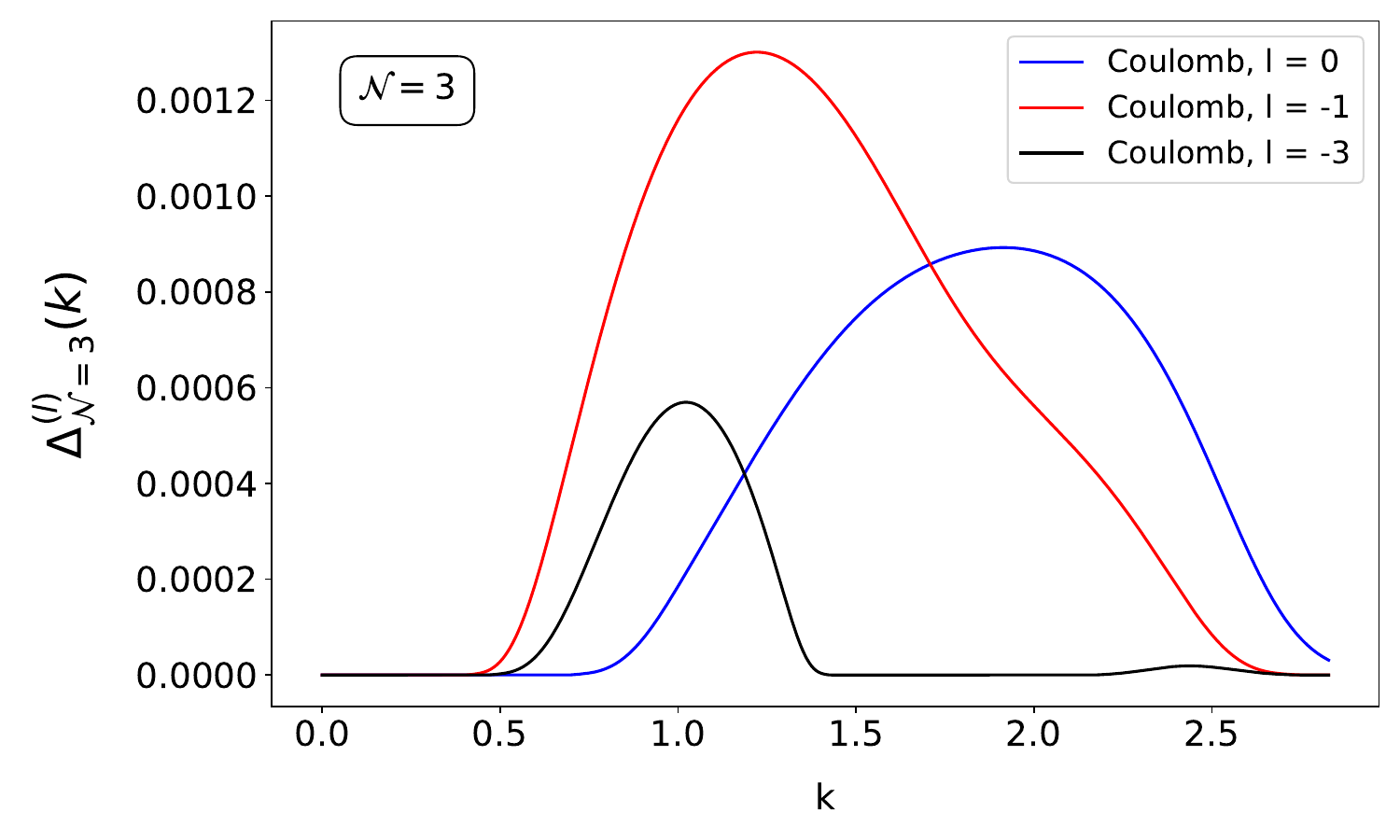}

\caption{The solutions for $ \Delta(\vec{k}) $ calculated self-consistently using Eq.~(\ref{selfConEq}) 
for different pairing channels $l$ in the lowest four LLs ($ \mathcal{N} = 0, 1, 2, 3 $) for the Coulomb interaction.}	
\label{selfConFig}
\end{figure*}

\begin{figure*}[ht!]
\centering
\includegraphics[width=0.48\linewidth]{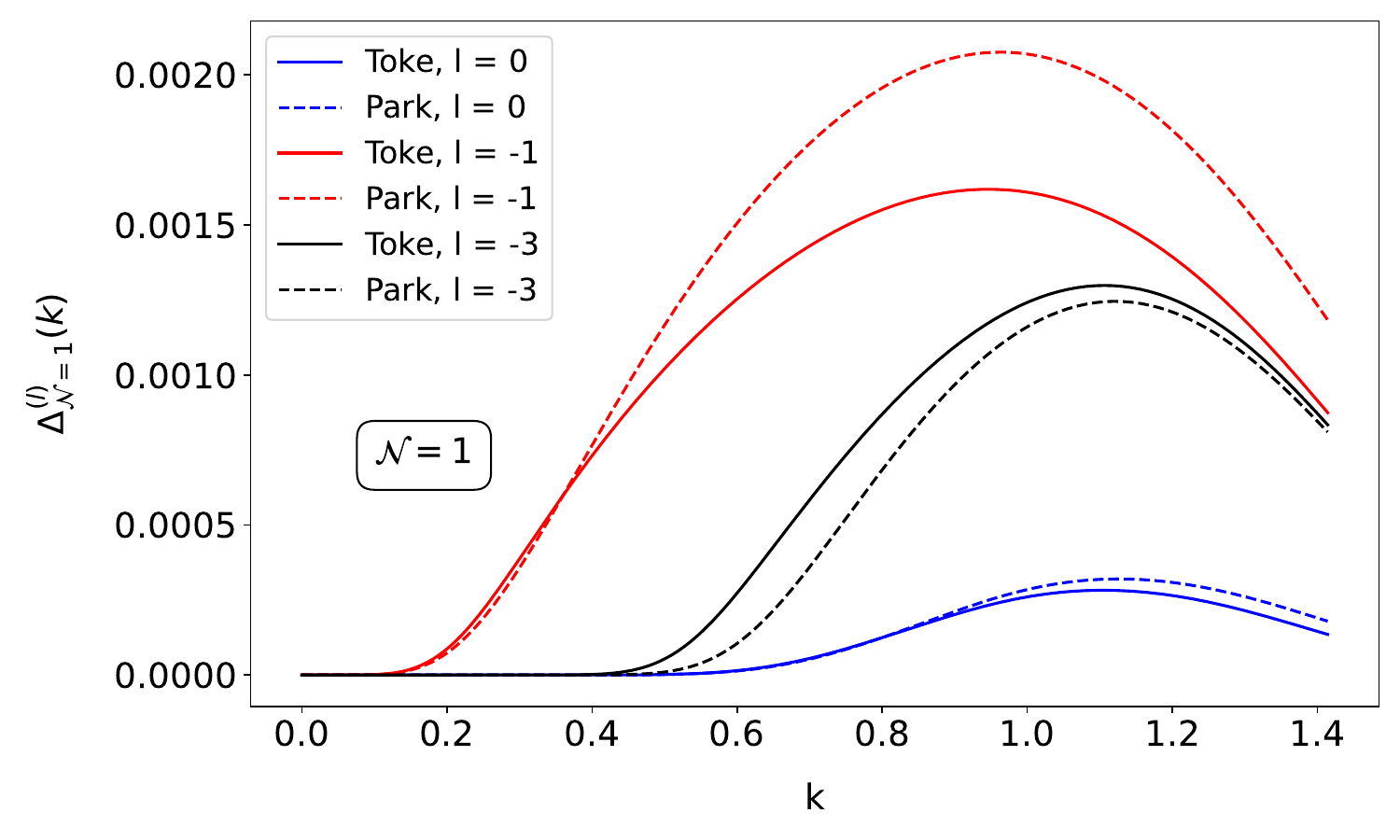}
\hfill
\includegraphics[width=0.48\linewidth]{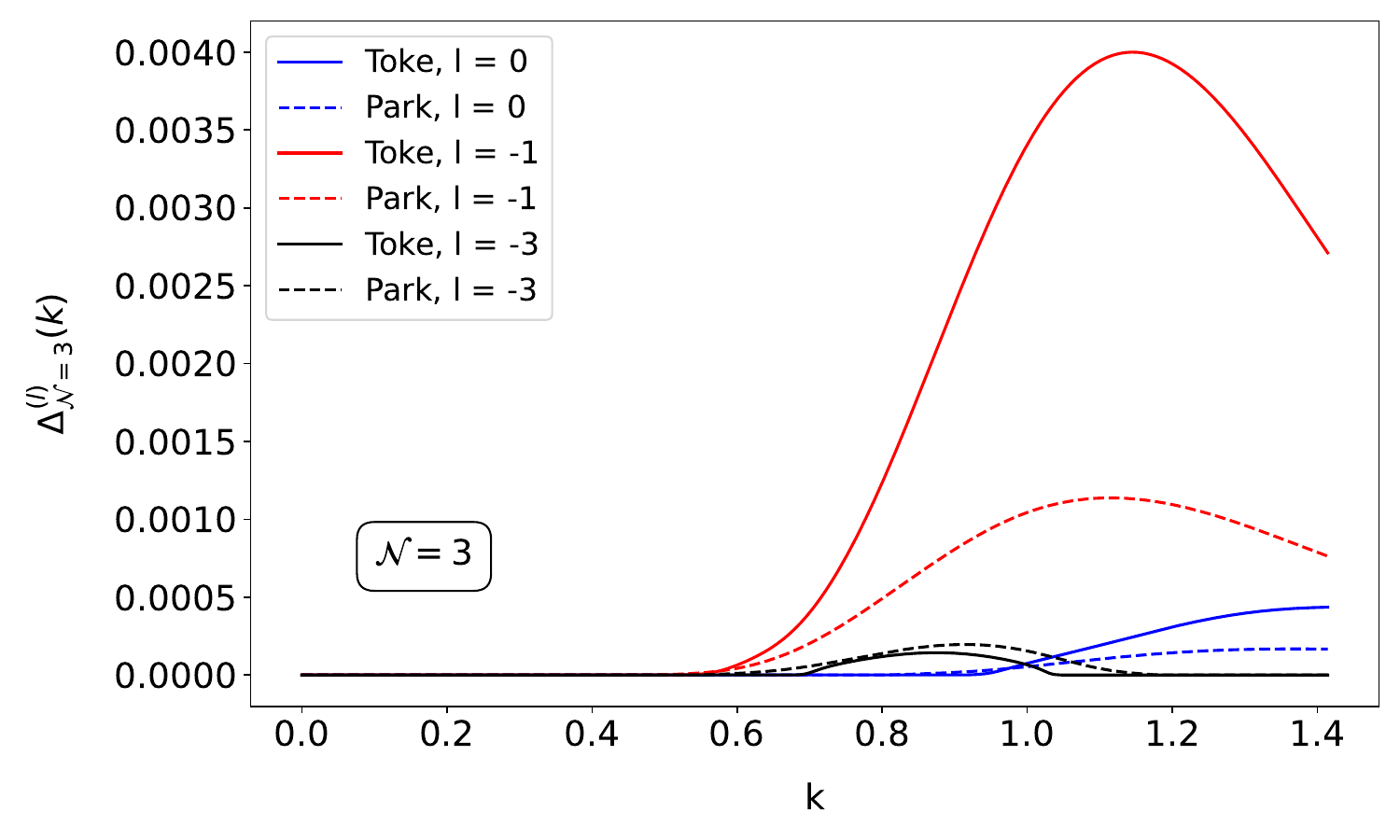}

\caption{
Self-consistently calculated solutions for $ \Delta(\vec{k}) $
for different pairing channels $l$ in the second ($ \mathcal{N} = 1 $) (left) and fourth ($ \mathcal{N} = 3 $) (right) Landau levels (LLs). These results correspond to the Toke and Park interactions.	}
\label{selfConFigHaldane}
\end{figure*}

As discussed earlier, a key feature of the Hamiltonian in the dipole representation of CFs is its inherent symmetry, particularly PH symmetry. This symmetry means that the system remains unchanged when particles are exchanged with real holes or when the densities $ \rho_L(\vec{q}) $ and $ \rho_R(\vec{q}) $ are swapped, where $ \rho_R(\vec{q}) $ corresponds to the density of correlation holes.

In the context of the Coulomb interaction, it is energetically favorable for these correlation holes to move away from the electrons, allowing real holes to surround the electrons instead. We have previously shown that the density of these correlation holes is equal to the density of real holes, and that the size of the resulting dipoles is determined by a translation operator. As a result, at the Fermi level $ k = k_F $, the electrons are positioned far from the holes, making this FLL state energetically unstable and prone to become a critical FLL state \cite{nikola}.

Breaking the symmetry between the left (L) and right (R) components of the system is equivalent to breaking the symmetry between particles and real holes. Therefore, states that break this symmetry are likely to be more energetically favorable. It is worth noting that at half-
filling, most of the paired states that emerge (such as the Moore-Read Pfaffian, anti-Pfaffian, and f-wave states) spontaneously break this symmetry.

Motivated by the aim to explore a well-defined state of CFs within the dipole representation, that 
stabilizes the critical state, we derive an effective Hamiltonian. This is achieved by revisiting the 
Hamiltonian in Eq.~(\ref{ham1}) and subsequently subtracting (or adding) the following term:
\small
\begin{equation}
\mathcal{H}_{1} = \frac{1}{8} \int \frac{d^2 \vec{q}}{(2\pi)^2} V^{(\mathcal{N})}(\vec{q}) \left(\rho_0^{R}(-\vec{q})+\rho_0^L(-\vec{q})\right)\left(\rho_0^{R}(\vec{q}) - \rho_0^{L}(\vec{q})\right).
\label{ham_zero2} 
\end{equation}
\normalsize
which effectively represents zero in the physical space. Consequently, the resulting effective Hamiltonians take the following forms:
\begin{equation}
\mathcal{H}_{\text{res}}^{(1)} = \frac{1}{4} \int \frac{d^2 \vec{q}}{(2\pi)^2} V^{(\mathcal{N})}(\vec{q}) \rho_0^{R}(-\vec{q})\left(\rho_0^{R}(\vec{q}) - \rho_0^{L}(\vec{q})\right),
\label{ham3} 
\end{equation}
and
\begin{equation}
\mathcal{H}_{\text{res}}^{(2)} = \frac{1}{4} \int \frac{d^2 \vec{q}}{(2\pi)^2} V^{(\mathcal{N})}(\vec{q}) \rho_0^{L}(-\vec{q})\left(\rho_0^{L}(\vec{q}) - \rho_0^{R}(\vec{q})\right).
\label{ham4} 
\end{equation}
The dipole representation of these Hamiltonians plays a crucial role in defining a single energy, which is essential for obtaining 
paired solutions. Additionally, they can also be interpreted as symmetry-breaking modifications of the 
Hamiltonian in Eq.~(\ref{ham1}), stabilizing the system into one of two paired states \cite{nikola}.

To explore potential paired solutions, we apply the HF approach to the relevant part of 
the Hamiltonian in Eq.~(\ref{ham3}), yielding:
\begin{small}
\begin{align}
\mathcal{H}_{HF}^{(\mathcal{N})} & = \int \frac{d^2 \vec{k}}{(2 \pi)^2} \, \xi_{\mathcal{N}}(\vec{k}) \,  c_{\vec{k}}
^{\dagger}c_{\vec{k}} \, + \nn \\
& + \frac{1}{4} \int \frac{d^2 \vec{q}}{(2 \pi)^2} \int \frac{d^2 \vec{k_1}}{(2\pi)^2} \int \frac{d^2 
\vec{k_2}}{(2\pi)^2}  \upsilon(\vert \vec{q} \vert) F_{\mathcal{N}}^{2}(\vert \vec{q} \vert) \left(1 - e^{i\vec{k}\times 
\vec{q}}\right) \nn \\ 
& \cdot \left(\langle c_{\vec{k}_1 + \vec{q}}^{\dagger}c_{\vec{k}_2-\vec{q}}^{\dagger} \, \rangle \, 
c_{\vec{k}_1}c_{\vec{k}_2} + c_{\vec{k}_1 + \vec{q}}^{\dagger}c_{\vec{k}_2-\vec{q}}^{\dagger} \, \langle 
c_{\vec{k}_1}c_{\vec{k}_2} \rangle - \right. \nn \\
& - \left. \langle c_{\vec{k}_1 + \vec{q}}^{\dagger}c_{\vec{k}_2-\vec{q}}^{\dagger} \, \rangle \, 
\langle c_{\vec{k}_1}c_{\vec{k}_2} \rangle \right),
\end{align}
\end{small}
where $ \xi_{\mathcal{N}}(\vec{k}) = \varepsilon_{\mathcal{N}}(\vec{k}) - \varepsilon_{\mathcal{N}}(\vec{k}_F)  $, with 
$ \epsilon_{\mathcal{N}}(\vec{k}) $ representing the dispersion relation given by:
\begin{align}
\varepsilon_{\mathcal{N}}(\vec{k}) & = \frac{1}{4} \int \frac{d^2 \vec{q}}{(2  \,\pi)^2} 
\upsilon(\vert \vec{q} \vert) F_{\mathcal{N}}^2(\vert \vec{q} \vert) \left(1 - e^{i \vec{k}\times \vec{q}}\right) \nn \\
& - \frac{1}{2} \int \frac{d^2 \vec{q}}{(2 \, \pi)^2} 
\upsilon(\vert \vec{k} - \vec{q} \vert) F_{\mathcal{N}}^2(\vert \vec{k} - \vec{q} \vert) \left(1 - e^{i \vec{k}\times \vec{q}}\right).
\end{align}
Furthermore, we define:
\begin{small}
\begin{align}
\Delta_{\mathcal{N}}(\vec{k})  = & - \frac{1}{4} \int \frac{d^2 \vec{q}}{(2 \pi)^2} \upsilon(\vert \vec{k} - \vec{q} \vert) F_{\mathcal{N}}^2(\vert \vec{k} - \vec{q} \vert) \left(1 - e^{-i \vec{k}\times \vec{q}}\right) \nn \\
& \cdot \langle c_{\vec{k}_1}c_{\vec{k}_2} \, \rangle
\end{align}
\end{small}
Finally, using standard BCS transformations to diagonalize the Hamiltonian in Eq.~(\ref{ham3}), we obtain the total energy:
\begin{equation}
E_{\text{paired}}^{(\mathcal{N})} = \int \frac{d^2 \vec{k}}{(2\, \pi)^2}\left(\xi_{\mathcal{N}}(\vec{k})-E_{\mathcal{N}}(\vec{k})\right) + 
\int \frac{d^2 \vec{k}}{(2\, \pi)^2} \frac{\vert \Delta_{\mathcal{N}}(\vec{k}) \vert^2}{2 E_{\mathcal{N}}(\vec{k})},
\end{equation}
where the Bogoliubov quasiparticle energy is given by $ E_{\mathcal{N}}(\vec{k}) = \sqrt{\xi_{\mathcal{N}}^2(\vec{k})+\vert 
\Delta_{\mathcal{N}}(\vec{k})\vert^2} $. In the BCS treatment, we obtain the pairing instability is described by 
the order parameter $ \Delta(\vec{k}) $ by self-consistently:
\begin{footnotesize}
\begin{equation}
\vert \Delta_{\mathcal{N}}^{(l)}(\vec{k}) \vert = \frac{1}{4} \int \frac{d^2 \vec{q}}{(2 \pi)^2} \upsilon(\vert \vec{k} - \vec{q} \vert) F_{\mathcal{N}}^2(\vert \vec{k} - \vec{q} \vert) \left(1 - e^{-i \vec{k}\times \vec{q}}\right) 
\frac{ \Delta_{\mathcal{N}}^{(l)}(\vec{q})}{2E_{\mathcal{N}}(\vec{q})}.
\label{selfConEq}
\end{equation}
\end{footnotesize}
Here, the gap function takes the following form: $ \Delta^{(l)}(\vec{k}) = e^{i l \theta_l} \vert 
\Delta(\vec{k})\vert$, where 
$ \theta $ is the angular coordinate of $ k $, with $ l = 0, \pm 1, \pm 3 $ representing 
different pairing channels.
Details regarding the numerical solutions of the self-consistent equation in Eq.~(\ref{selfConEq}) are 
provided in the Appendix \ref{appendixB}. We note that we obtain the same solutions for $ l = 1 $ in 
Hamiltonian Eq.~(\ref{ham4}) as we do for $ l = -1 $ in Hamiltonian Eq.~(\ref{ham3}).

\begin{figure}[h!]
    \includegraphics[width=0.9\linewidth]{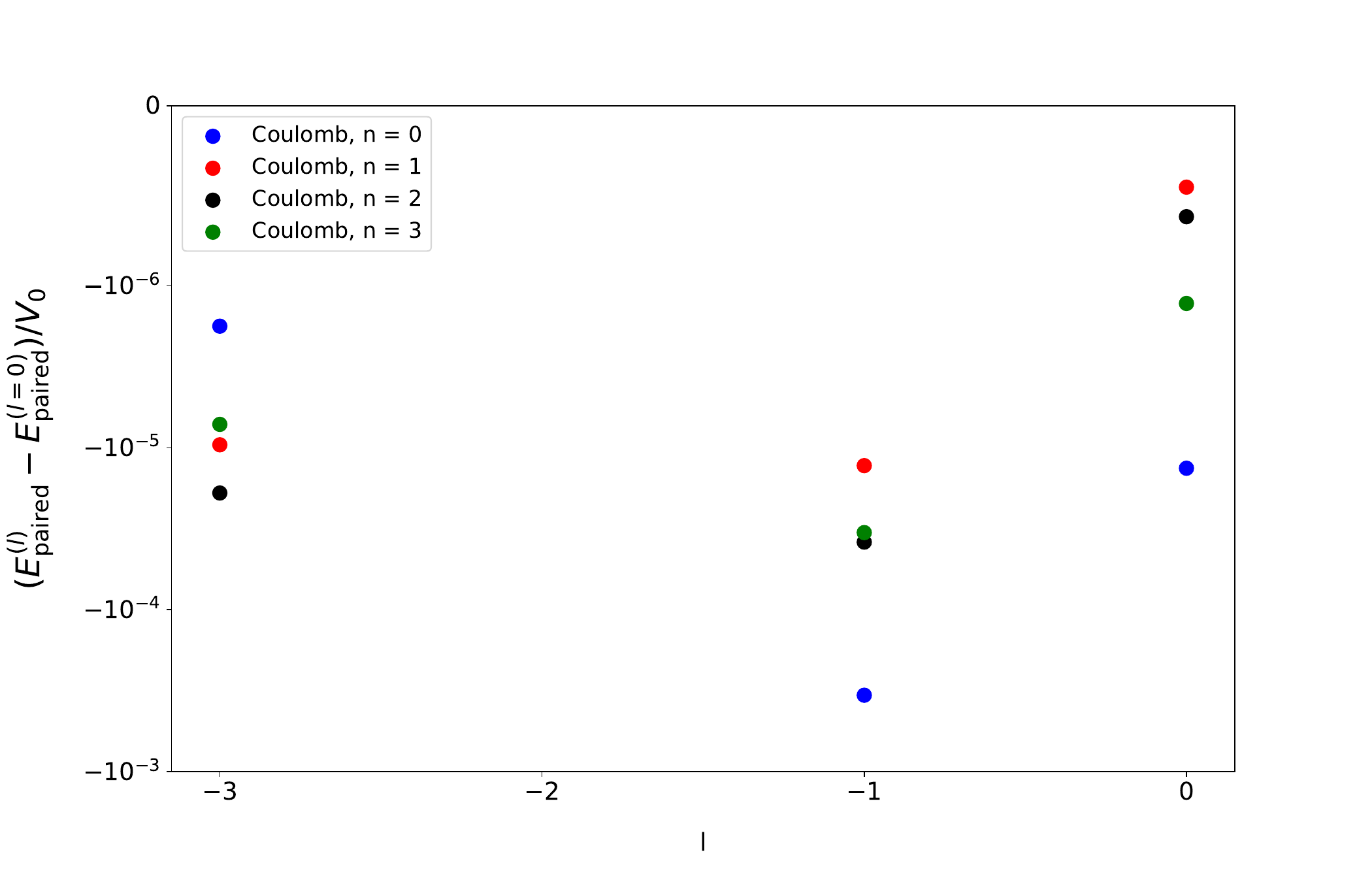}\\
    \includegraphics[width=0.9\linewidth]{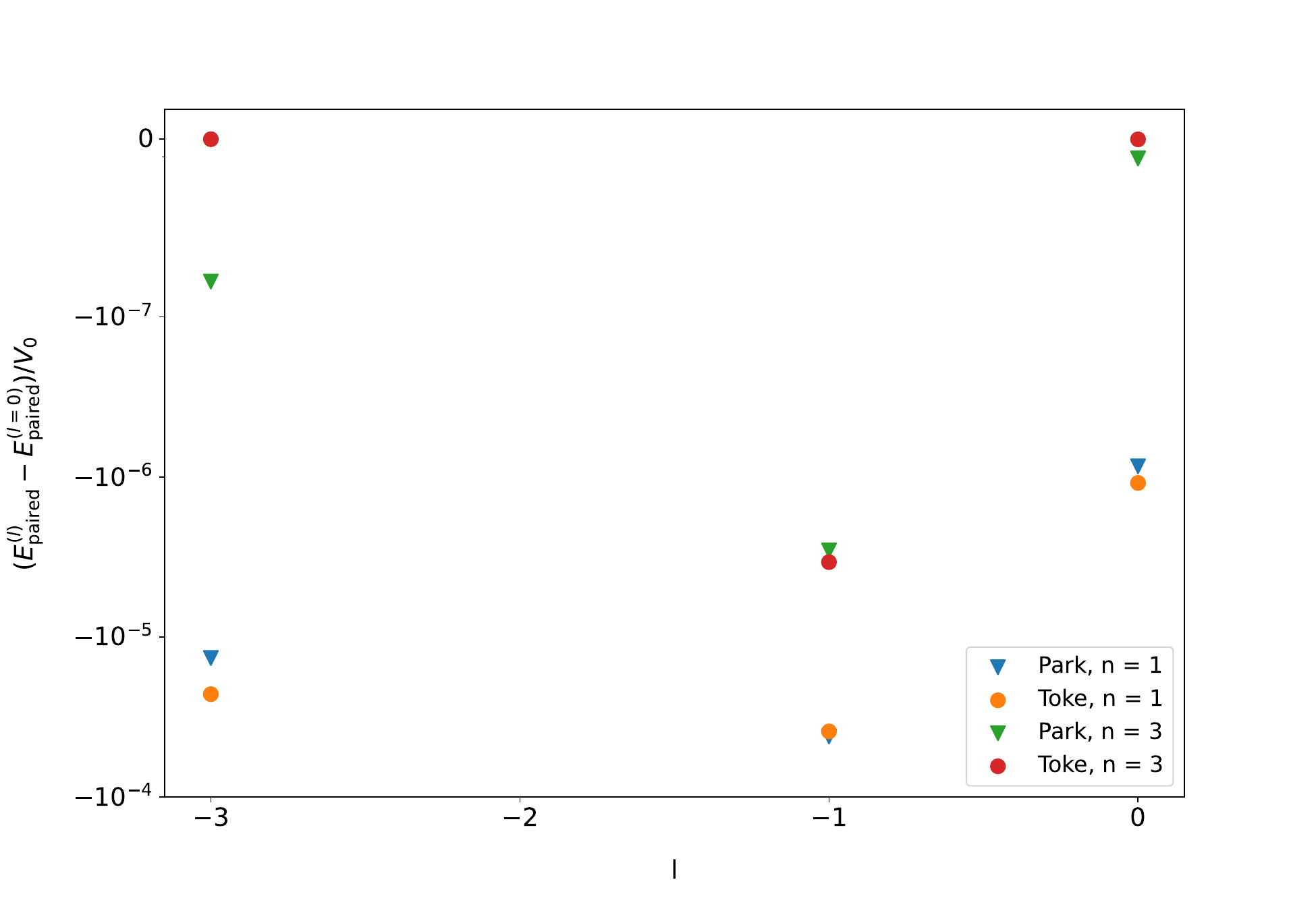}
    \caption{Comparison of the total energy for various pairing solutions, expressed as $ E_{\text{paired}}(l)-E_{\text{paired}}(l=0)  $ for Coulomb interactions (upper), and Toke and Park interactions (lower). The interaction strength is given by $V_0 = 2 \pi e^2 / l_B$.}
    \label{Epaired}
\end{figure}

In the previous section, we regularized the state of the Hamiltonian in Eq.~(\ref{ham_res}) by eliminating the 
quadratic term, which contributes to the kinetic energy. Consequently, the regularized state cannot 
support gapped pairing solutions due to the absence of (bare) mass. While the critical state can be 
stabilized among the paired states, it can also be stabilized within the regularized state. To 
investigate in which of these states the critical state is stabilized, we define the energy of the 
regularized state of CFs in the dipole representation using a mean-field approach as follows:

\begin{equation}
E^{(\mathcal{N})}_{\text{FLL}} = \int \frac{d^{2}\vec{k}}{(2\pi)^2} \left(\varepsilon^{(\mathcal{N})}_{\text{0}}(\vec{k}) + \frac{1}{2}\varepsilon^{(\mathcal{N})}_{\text{HF}}(\vec{k})\right).
\end{equation}

In calculating the total energy, we apply a "cut-off" when necessary, using the radius of a circle in 
momentum space $ Q = \sqrt{2}\, k_F $, which denotes the volume of available states in the LL. 
Furthermore, the only viable cut-off value for the regulized FLL state being $ k_F $, as it is designed 
to accurately describe the physics at the Fermi surface $ k_F $. We have carefully used the cut-off \( Q = \sqrt{2} \, k_F\) for the paired state, but we have also ensured that our findings do not depend on the choice of the cut-off value. In spherical coordinates, this 
corresponds to $ Q = \sqrt{2} \, k_F $. 
\\

\begin{figure}[h!]
    \includegraphics[width=0.9\linewidth]{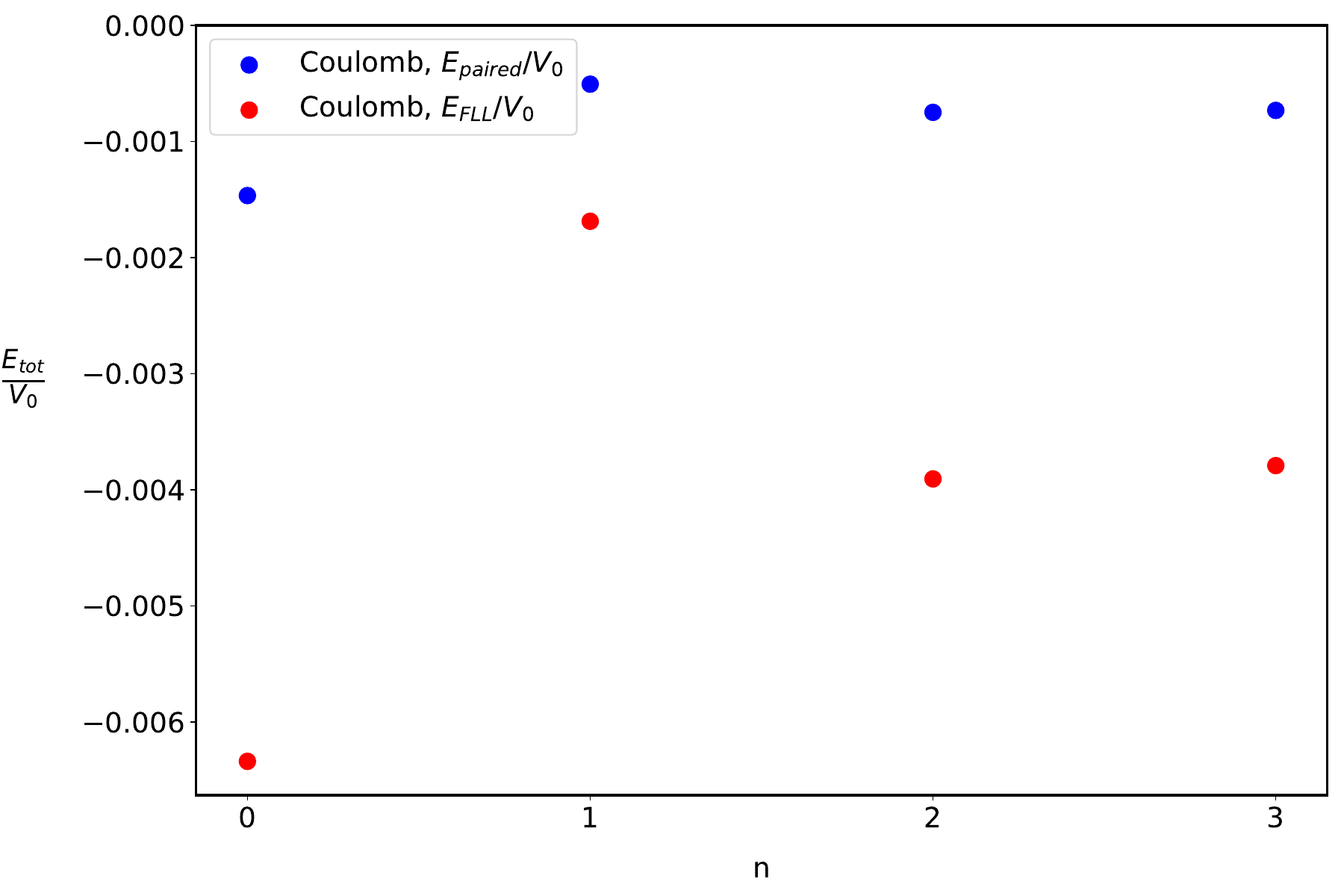}\\
    \includegraphics[width=0.9\linewidth]{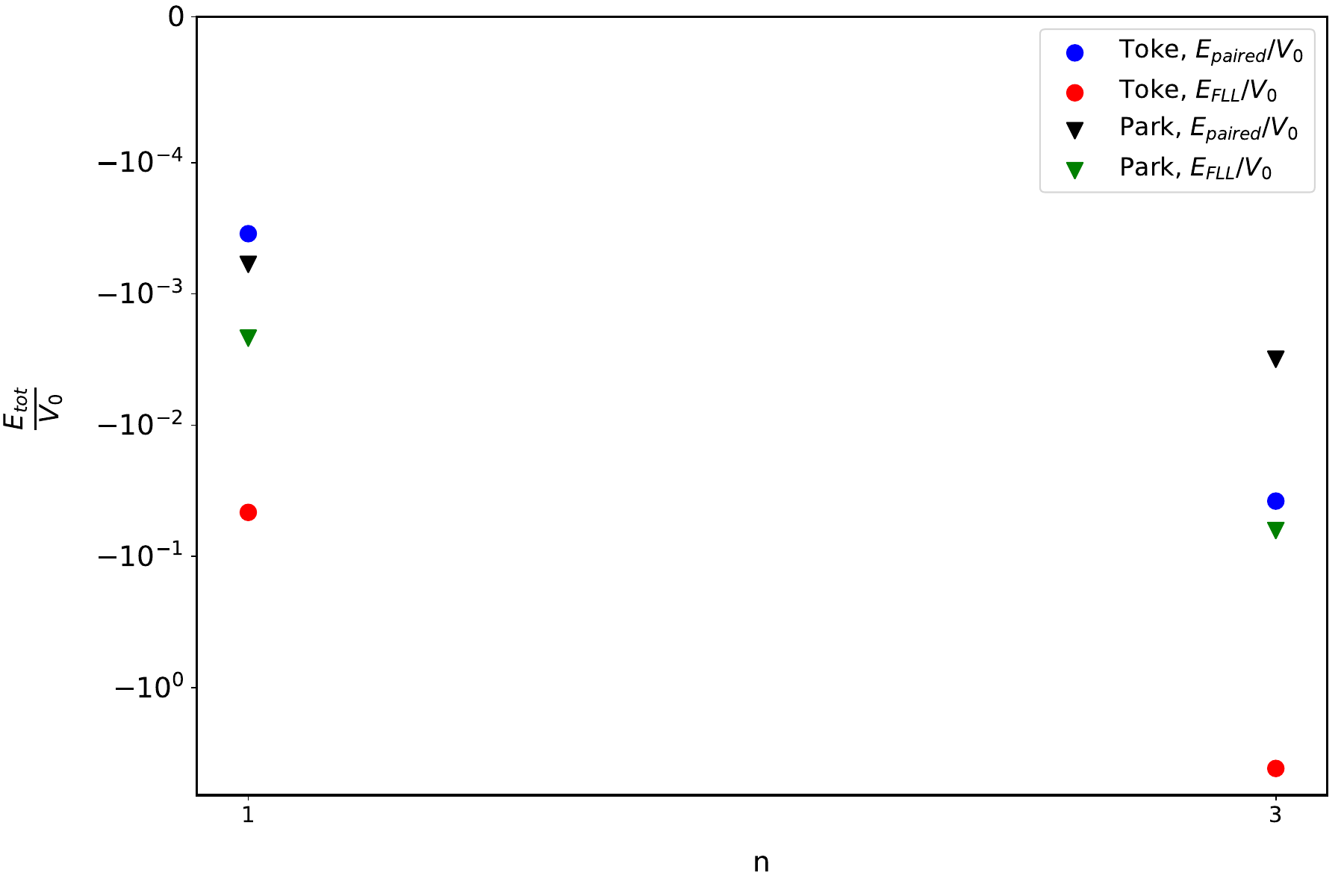}
    \caption{Comparing of total energies of the pairing solutions $ E_{\text{paired}} $ and the regularized state $ E_{\text{FLL}} $ in the four lowest LLs in the case of Coulomb interactions (upper), Toke, and Park interactions (lower). Here, we calculated $ E_{\text{paired}} $ for $ l = -1 $, which minimizes the energy value and is therefore the most energetically favorable. The interaction strength is defined by $V_0 = 4 \pi e^2 / l_B$.}
    \label{Etot}
\end{figure}

We illustrate in Fig.~(\ref{selfConFig}) the self-consistently obtained mean-field 
parameter $ \Delta(\vec{k}) $ in the fourth LL $ \mathcal{N} = 0, 1, 2, 3 $ for the Coulomb potential, 
and the Fig.~(\ref{selfConFigHaldane}) for the Toke and Park potentials. Here, it can be notice that non-zero pairing solutions 
also appear in the LLL, which might suggest that pairing could persist in the thermodynamic limit. 
However, this stands in contrast to numerical studies, which consistently find the FLL state to be the 
most stable configuration. The definitive confirmation about the presence or absence of pairing lies on 
a direct comparison of the total energies between a paired state and a regularized FLL state. Additionally, we plotted in the Fig.~(\ref{Epaired}) the total energy $ E_\text{paired} $ of the different pairing solutions compared to the normal state energy to determine the energetically most favorable $ l $ in the case of Columb and Haldane potentials.
In Fig.~(\ref{Etot}), we compare the results for $ E_{\text{FLL}} $ and $ E_{\text{paired}} $ in the case of graphene for the four lowest LLs ($ \mathcal{N} = 0, 1, 2, 3 $) for the Coulomb, Toke and Park interactions to investigate the possibility of pairing 
instability. We would like to note that here we calculated $ E_{\text{paired}} $ to compare the energy for the self-consistent solution that minimizes the energy value. We find that the critical state is stabilized within the regularized state, which is 
energetically more favorable than the paired states. In the following sections, we will discuss these 
findings in more detail for each of the four lowest LLs ($ \mathcal{N} = 0, 1, 2, 3 $), emphasizing how our results align with previous 
experiments and numerical studies. \\

In the lowest LL ($ \mathcal{N} = 0 $), considering the short-range, repulsive Coulomb interaction 
$ \mathcal{V}(|\vec{q}|) = q^2 $, which describes a two-body interaction between dipoles, 
the three-body Hamiltonian can be derived using the Chern-Simons approach \cite{geiter1982, geiter1992} 
as follows:
\begin{equation}
\mathcal{H}_{\text{eff}}^{(e)} = \sum_{\langle ijk \rangle} \nabla^2_i \delta^2(\vec{x}_i - \vec{x}_j) 
\delta^2(\vec{x}_i - \vec{x}_k),
\label{three-body}
\end{equation}
which aligns with the commonly used interaction model for the Pfaffian \cite{moore}. In the lowest LL
($ \mathcal{N} = 0 $), 
the effective interaction between CFs (and therefore the physics in general) in 
monolayer graphene is the same as in a 2DES, such as GaAs quantum wells with zero width. Here, numerical experiments indicate that the Fermi-liquid-like state is the most stable in the half-
filled first LL ($ \mathcal{N} = 0 $). \\

In the half-filled second LL ($ n = 1 $), non-trivial topology is identified via Majorana edge states in the case of 
2DES. In this LL, well-defined $ p $-wave solutions in 2DES are present, particularly 
$ l = 1 $ ($ l = -1 $) wave pairing in Hamiltonian as described in Eq.~(\ref{ham4}) (Eq.~(\ref{ham3})), but with 
form factors characteristic of 2DES \cite{nikola}. This corresponds to the Pfaffian topological order 
\cite{moore}. The PH conjugate of the Pfaffian, the anti-Pfaffian, can be obtained by 
switching the sign of the particles \cite{levin, lee2007}. We would like to point out here that a well-
defined paired state can only exist if the effective dipole physics at the Fermi level, driven by the 
non-trivial topology of the LL, is present. However, in monolayer graphene, in the second LL ($\mathcal{N} = 1$) the regularized state 
dominates over the paired state, which is consistent with earlier studies indicating that the lowest 
energy state is a Fermi sea of CFs with no pairing instabilities \cite{balram2015}. \\

In the half-filled third LL ($ \mathcal{N} = 2 $) of graphene, the FQHE has not been observed 
\cite{diankov, chen}.
However, in numerical experiments on a torus using a BCS variational state for CFs, the authors in
Ref. \cite{sharma} identified the presence of pairing instability, which most likely 
corresponds to $ p $-wave pairing. 
This may result in the observation of pairing solutions in the thermodynamic limit, which is well represented in numerical experiments on a torus. In 
contrast with this numerical study constructed on torus, we propose that the regularized state is energetically more 
favorable, which does not exhibit well-defined pairing solutions. 
\\

Finally, in previously mentioned study \cite{sharma}, it has been proposed that the Fermi sea of CFs may 
be unstable to $ f $-wave pairing in the half-filled fourth LL ($ \mathcal{N} = 3 $) of graphene. 
This particular LL is significant due to the experimental observation of the FQHE \cite{kim2019}. 
However, numerical studies conducted on a spherical geometry reveal that neither the anti-Pfaffian nor 
the PH symmetric Pfaffian \cite{mishmash} states exhibit a strong overlap with the exact ground state 
near the pure Coulomb interaction point. 
Additionally, various spin and valley singlet states were examined, yet none showed significant overlap 
with the ground state in the $\mathcal{N} = 3$ LL. Although the 221-parton 
state \cite{jain1989} demonstrates a notable similarity to the pure Coulomb interaction across a wide 
parameter range, its validation still requires further theoretical and experimental studies. Our 
findings suggest that the critical state in the fourth LL ($ \mathcal{N} = 3 $) is not stabilized among the paired 
states, such as the Pfaffian. 
Instead, the regularized state, which lacks well-defined pairing 
instabilities, emerges as energetically more favorable. This suggests the absence of well-defined 
pairing instabilities within the dipole representation of CFs in this context.  Therefore, our results 
are consistent with numerical studies conducted on spherical geometry. To conclusively validate the 
nature of the ground state and fully elucidate the pairing mechanism, particularly in higher LLs, 
additional theoretical and experimental research extending beyond the framework of CFs is essential. 
Strong short-range repulsive interaction is crucial for defining CFs in lower LLs, where the effects of 
this interaction are dominant (see Eq.~(\ref{three-body})).

\section{Conclusions}
\label{conclusion}

In this study, we have derived the dipole representation of CFs in graphene’s quantum Hall systems, focusing particularly on the half-filled LLs. Our investigation extended the Pasquier-Haldane-Read construction to describe the excitations in half-filled LLs of electrons in graphene, and we derived an effective Hamiltonian within the dipole framework that respects the key symmetry of these systems, including PH symmetry. We demonstrated that this symmetry, inherent in the effective Hamiltonian, can be broken and stabilized either in the phase of paired states or in a regularized state without bare mass at the Fermi level.

We discussed the relationship of paired states to the Pfaffian state. Furthermore, our analysis of the pairing mechanism shows that the regularized state, which lacks well-defined pairing instabilities, emerges as the energetically favored state over the paired states with p-wave and f-wave pairing. This finding suggests that, within the dipole representation, the ground state in the half-filled fourth LL ($\mathcal{N} = 3$) of graphene is not characterized by pairing instabilities.

We also discussed the consistency of our results with experimental and numerical studies. 
Furthermore, we highlighted the limitations of the dipole representation framework.
Additionally, going beyond the mean-field theory and including fluctuations might be crucial in accurately describing the true nature of the ground state.

We hope this study will inspire further investigations of these states, including the pairing mechanism and the origin of the ground state. Notably, the experimental observation of FQHE in bilayer graphene systems \cite{papic} suggests that similar approaches could be applied to understand these phenomena in other layered or structured graphene systems.

\begin{acknowledgments} The autor is very greatful to Milica Milovanovi\'c to previous work on 
related topics. S.P. kindly thanks Ajit Balram for useful discussions, and Antun Bala{\v z}, Nenad Vukomirovi\'c, and Jak{\v s}a Vu{\v c}i{\v c}evi\'c for helful advice. Computations were performed on the PARADOX supercomputing facility (Scientific 
Computing Laboratory, Center for the Study of Complex Systems, Institute of Physics Belgrade). The 
author acknowledges funding provided by the Institute of Physics Belgrade, through the grant by the 
Ministry of Science, Technological Development, and Innovations of the Republic of Serbia. Furthermore, 
S.P. acknowledges funding supported as returning expert by the Deutsche Gesellschaft für Internationale 
Zusammenarbeit (GIZ) on behalf of the German Federal Ministry for Economic Cooperation and Development 
(BMZ). 
\end{acknowledgments}

\appendix
\input{AppendixA}

\input{AppendixB}



\end{document}

%% file: AppendixA.tex
\section{Analytical Derivation of Single-Particle Energies for the Lowest Four Landau Levels}
\label{appendixA}
The Hamiltonian in the dipole representation, as described by Eq.~(\ref{ham1}), is given by:
\begin{small}
\begin{equation}
\mathcal{H}_{\text{eff}} = \frac{1}{8} \int \frac{d^2 \vec{q}}{(2\pi)^2} V^{(\mathcal{N})}(\vec{q}) \left(\rho_{L}(\vec{q}) - \rho_{R}(\vec{q})\right)\left(\rho_{L}(-\vec{q}) - \rho_{R}(-\vec{q})\right).
\end{equation}
\end{small}
The energy dispersion relation for graphene in the $\mathcal{N}$-th Landau level (LL) is expressed as:
\begin{equation}
\varepsilon(\vec{k}) = \varepsilon_{0}(\vec{k}) + \varepsilon_{HF}(\vec{k}),
\label{energyTotal}
\end{equation}
where:
\begin{equation}
\varepsilon_{0}^{(\mathcal{N})}(\vec{k}) = \frac{1}{2}\int \frac{d^2 \vec{q}}{(2\pi)^2}V^{(\mathcal{N})}(\vert \vec{q} \vert)\sin^2\left(\frac{\vec{k}\times \vec{q}}{2}\right),
\label{energyTotal0}
\end{equation}
represents the single-particle energy, and
\begin{equation}
\varepsilon_{HF}^{(\mathcal{N})}(\vec{k}) = - \int \frac{d^2 \vec{q}}{(2\pi)^2}V^{(\mathcal{N})}(\vert\vec{k}-\vec{q}\vert)\sin^2\left(\frac{\vec{k}\times \vec{q}}{2}\right)n_{q},
\label{energyTotalHF}
\end{equation}
represents the Hartree-Fock (HF) contributions.

In this Appendix, we analytically derive the single-particle energies for the first four Landau levels. We begin by utilizing the identity $ \cos(a) = \frac{1}{2}\left(e^{i a} + e^{-i a}\right) $, which allows us to express the single-particle energy in $ \mathcal{N} = 0 $ as:
\begin{align}
\varepsilon^{(0)}_{0}(k)=&\int^{\infty}_0\frac{dq}{8\pi}e^{-\frac{q^2}{2}}\nn \\
-&\int^{\infty}_0
\int^{2\pi}_0\frac{dq d\phi}{32\pi^{2}}e^{-\frac{q^2}{2}}\left(e^{ikq\sin(\phi)}+e^{-ikq\sin(\phi)}\right).
\end{align}

The equality $ \text{erf}(x) = \frac{2}{\sqrt{\pi}}\int_0^{x}e^{-t^2}dt $ motivates us to introduce the substitution $ u = \frac{q}{\sqrt{2}} $, which transforms the expression into:
\begin{align}
\varepsilon^{(0)}_{0}(k)=&\int_0^{\infty}\frac{du}{8\sqrt{2}\sqrt{\pi}}\frac{2}{\sqrt{\pi}} e^{-u^2} \nn \\
-&\int_0^{\infty}\int_0^{2\pi}\frac{dq d\phi}{32\pi^{2}}e^{-\frac{1}{2}k^2 \sin^2(\phi)}\left(e^{-\frac{1}{2}(q-i k \sin(\phi))^2}\right. \nn \\
+&\left. e^{-\frac{1}{2}(q+ik\sin(\phi))^2}\right).
\end{align}

Next, by introducing the substitutions $ v_1 = \frac{1}{\sqrt{2}}(q-ik\sin(\phi)) $ and $ v_2 = \frac{1}{\sqrt{2}}(q+ik\sin(\phi)) $, we obtain:
\begin{align}
\epsilon^{(0)}_{0}(k) = \frac{1}{8\sqrt{2}\sqrt{\pi}} - \int_0^{2\pi}\frac{d\phi}{16\sqrt{2}\pi\sqrt{\pi}}e^{-\frac{1}{2}k^2\sin^2(\phi)}.
\end{align}

Finally, utilizing the identity $ \sin^{2}(\phi) = 1 - \cos^{2}(\phi) $, we derive the final expression for the single-particle energy:
\begin{align}
\epsilon_0^{(0)}(k) = \frac{1}{8\sqrt{2}\sqrt{\pi}}\left(1-e^{-\frac{1}{4}k^2}I_0\left(\frac{k^2}{4}\right)\right),
\end{align}
where $ I_0(k) $ is the modified Bessel function of the first kind.

Analogously, we obtain the single-particle energies for the other LLs:
\begin{widetext}
\begin{align}
\varepsilon_0^{(1)}(k) = & \frac{1}{256 \sqrt{2 \pi}} \left(22 - e^{-\frac{k^2}{4}} \left( \left(22 + 2 k^2 + k^4\right) I_0\left(\frac{k^2}{4}\right) - k^2 \left(4 + k^2\right) I_1\left(\frac{k^2}{4}\right)\right) \right) \\ 
\varepsilon_0^{(2)}(k) = & \frac{1}{4096 \sqrt{2 \pi}} \left(290 - e^{-\frac{k^2}{4}} \left(\left(290 - 12 k^2 + 28 k^4 - 2 k^6 + k^8\right) I_0\left(\frac{k^2}{4}\right) 
 + k^2 \left(56 + 30 k^2 + k^6\right) I_1\left(\frac{k^2}{4}\right)\right)\right) \\
\varepsilon_0^{(3)}(k) = & \frac{1}{147456 \sqrt{2 \pi}} \left(9270 - e^{-\frac{k^2}{4}} \left(9270 - 1458 k^2 + 1809 k^4 - 360 k^6 + 114 k^8 - 14 k^{10} + k^{12}\right) I_0\left(\frac{k^2}{4}\right) \right. \nn \\
+ & e^{-\frac{k^2}{4}} \left. k^2 \left(1836 + 1563 k^2 - 192 k^4 + 92 k^6 - 12 k^8 + k^{10}\right) I_1\left(\frac{k^2}{4}\right)\right)
\end{align}
\end{widetext}

%% file: AppendixB.tex
\section{Numerical methods}
\label{appendixB}

In this section, we describe the numerical methods used to solve the self-consistent equation 
( Eq.~(\ref{selfConEq})) using the trapezoidal rule algorithm. 
The integral on the right-hand side of Eq.~(\ref{selfConEq}) is computed by discretizing the momentum space using a uniform grid. For each grid point, the integral is approximated by summing the contributions from neighboring points, weighted by the trapezoidal rule. The integration is performed iteratively, starting with an initial guess for the mean-field parameter $\Delta(\vec{k})$ (for example, $
\Delta(\vec{k})=10^{-5}$ for all $k$) and updating it until the convergence criterion is satisfied. The criterion for convergence is set as:

\begin{equation}
\frac{\mathrm{max}_k |\Delta^\mathrm{new}_k - \Delta^\mathrm{old}_k|}{\mathrm{max}_k |\Delta^\mathrm{new}_k|} < 10^{-3} 
\end{equation}

It typically takes between 11 and 70 iterations to meet the convergence criterion. \\

We perform the integration on the RHS of Eq.~(\ref{selfConEq}) using the trapezoidal rule.
The trapezoidal rule for a two-dimensional integral over a grid can be expressed as:
\begin{widetext}
\begin{eqnarray}
\int_a^b \int_c^d f(x, y) \, dx \, dy \approx 
\sum_{i=1}^{N_x-1} \sum_{j=1}^{N_y-1} \frac{\Delta x \, \Delta y}{4} \left[f(x_i, y_j) + f(x_{i+1}, y_j) +  f(x_i, y_{j+1}) + f(x_{i+1}, y_{j+1})\right]
\end{eqnarray}
\end{widetext}
where \(\Delta x\) and \(\Delta y\) are the grid spacings in the \(x\) and \(y\) directions, respectively, and \(N_x\) and \(N_y\) are the number of grid points in each direction.

This numerical approach ensures that the results are accurate and independent of the specific choice of numerical parameters, providing a robust solution to the self-consistent equation.